\documentclass[12pt]{article}
\usepackage{graphicx}
\usepackage{amsmath,amsfonts,amssymb}
\usepackage[normalem]{ulem}

\input{tcilatex}

\begin{document}

\title{A critical assessment on Kassapoglou's statistical model for
composites fatigue }
\author{M. Ciavarella \\
Politecnico di BARI, V.le Gentile 182, 70125\ Bari, Italy \\
Corresponding author: mciava@poliba.it \\
[10pt] V. Vinogradov \\
School of Civil Engineering \& Geosciences, \\
Newcastle University, Newcastle upon Tyne, NE1 7RU, U.K. \\
[10pt] G. Carbone \\
Politecnico di BARI, V.le Gentile 182, 70125\ Bari, Italy}
\date{}
\maketitle

\begin{abstract}
Kassapoglou recently proposed a model for fatigue of composite materials
which seems to suggest that fatigue\ SN curve can be entirely predicted on
the basis of the statistical distribution of static strengths.\ The original
abstract writes \textit{\textquotedblleft Expressions for the cycles to
failure as a function of R ratio are derived. These expressions do not
require any curve fitting and do not involve any experimentally determined
parameters. The fatigue predictions do not require any fatigue tests for
calibration"}. These surprisingly ambitious claims and attractive results
deserve careful scrutiny. We contend that the results seem to be due to a
number of approximations and incorrect derivations, and one particular
misleading assumption, which make the model not conform to a fatigue testing
in a given specimen with resulting SN curve distribution. The quantitative
agreement of some predictions (the scatter of distribution of fatigue lives
being close to the mode value found in typical composites of aeronautical
interest in the large Navy database) should not motivate any enthusiasm. It
is believed that a proper statistical treatment of the fatigue process
should \textit{not }make wear-out constants disappear, and hence the SN
curves would depend on them, and not just on scatter of static data. These
serious concerns explain the large discrepancies found by 3 independent
studies which tried to apply Kassapoglou's model to composite fatigue data,
and to other well known results.
\end{abstract}


\section*{\protect\small Keywords}

{\small Composite materials, fatigue, wearout models, Kassapoglou model,
strength-life equal rank, statistics, Weibull distribution }

\section{\protect\normalsize Introduction}

{\normalsize The \textquotedblleft strength-life equal rank
assumption\textquotedblright\ wear-out models for fatigue of composite
materials were first presented by Hahn and Kim [1], and later as a fitting
approach to fatigue data by Sendeckyj [2]. A significantly different model
has been proposed more recently by Kassapoglou [3,4,5] (in the following, K
is Kassapoglou), which in fact claims an extremely strong result, which
defies the efforts of more than a century of research in fatigue: \textit{%
that of predicting SN\ curve of a material from just the static data}. The
effort started already in the end of the 1800's for metals, and even today
only crude approximations can be made on fatigue limit over static strength\
(the so-called, \textit{fatigue ratio}), which are generally based on
hardness tests. The use of simplified equations for SN curve also is well
known in any fatigue textbook [6], but it is always clearly shown that any
such empirical equation is limited, and in general makes only a \textit{very
crude} estimate. This justifies the industry of fatigue machine testing,
which is by no means less flourishing in composite materials, although
composites are known to suffer more crucially to impact than to fatigue.
Hence, no aircraft flying today is there without having passed a very
serious fatigue testing certification procedure, and for a good reason.
Fatigue testing is required by any certification agency to get airwhortiness
certificates, and the cost of testing is huge. The idea to obtain even
approximate results for fatigue from just static data is therefore still
obviously attractive, since experience on static strength is so much easier
and cheaper to obtain. Therefore, it is surprising to read in K's theory [3]
that \textit{\textquotedblleft Expressions for the cycles to failure as a
function of R ratio are derived. These expressions do not require any curve
fitting and do not involve any experimentally determined parameters. The
fatigue predictions do not require any fatigue tests for calibration"}.
Further, that \textit{\textquotedblleft comparison to several test cases
found in the Literature show this first simple model to be very promising", }
where for \textquotedblleft several test cases", K intends a few references
(references [34-41] in his paper [3], which are [7-14] here), where the
error is said to be small, but which in fact is not necessarily so. Take
Fig.6 of K's paper [3], where the agreement is said to be \textquotedblleft
very good": K's curve, which should be the median value, is seen to pass
close to the lowest data, and hence the error in terms of life can be easily
of 2 orders of magnitude.\ It is surprising that, after so much emphasis to
statistical treatment, the assessment of the quality of data fit is so poor,
and so much in favour of his own proposal. Not much better can be said
regarding Fig.9 (where the author admits the \textquotedblleft agreement to
be not so good"), obviously the author prefers to measure the error based on
stress, and claims a 17\% error is found -- it is not clear what kind of
crude estimate he is making given the so few data, but clearly the error in
terms of live can be various orders of magnitude. Similar problems were
found in Fig.10, and Fig.11, where the data are so few that do not really
deserve much attention. }

{\normalsize Therefore the sentence in the conclusion \textquotedblleft 
\textit{The approach allows analytical determination of the ratio of mean to
B- or A-basis life which can be used in designing certification of
qualification programs}" seems premature. The data show that the link with
static scatter is not as strong as to make any remote estimates of the SN
curve slopes. }

{\normalsize More recent assessment of K's model confirm this much more
cautious view, and indeed find naif that anyone could expect much from
static data statistical analysis, or when using K's method as
\textquotedblleft predictive" as such, warn of the surprisingly poor results
[15-17]. We shall draw attention in this paper to the fact that we do not
expect theoretically any reason for a good predictive capability to be
realistic in general. }

{\normalsize Certainly, there is some connection between static data and
fatigue limit (and therefore the SN\ curve slope tend to be similar for
similar materials), but the fact that the SN curve slope should depend
purely on the scatter of static data is an interesting and very potent
result, if only it were true. This happens to be what predicted by K's
model, particularly simple in the form assuming a Weibull distribution of
static strengths in which case the distribution of fatigue life is always an
exponential one, independent on either strength or scatter of strength of
the base material. The tendency to exponential distribution of fatigue
lives, which was observed before [8, 18] also in much larger set of data,
such a Navy database, is not a result of K's assumption. }

\section{\protect\normalsize A view on firm evidence from large databases}

{\normalsize Fleck Kang and Ashby [18], in an authoritative review which
contains also data on composite materials, produce a large set of maps
covering a huge number of references, and in particular show in Figure \ref%
{fig:Fleck} the well-known fact that the endurance limit $\sigma _{e}$
scales in a roughly linear way with the yield strength, $\sigma _{y}$.The
fatigue ratio, defined as $\sigma _{e}$/$\sigma _{y}$ (but more classically
for metals, $\sigma _{e} $/$\sigma _{fs}$) at load ratio $R=-1$, appears as
a set of diagonal contours. \textquotedblleft \emph{The value of fatigue
ratio, for engineering materials, usually lies between 0.3 and 1. Generally
speaking, it is near 1 for monolithic ceramics, about 0.5 for metals and
elastomers, and about 0.3 for polymers, foams and wood; the values for
composites vary more widely--from 0.1 to 0.5}\textquotedblright . Naturally,
for fatigue limit in composites (as well as light alloys), it is often
intended the value at a given fixed number of cycles. This wide variation
already makes one wonder that for composites the fatigue properties depend
less on static properties than for other materials. A first alarm bells
rings, with respect to Kassapoglou's claim. Moreover, Fleck Kang and Ashby
[17] remark \emph{\textquotedblleft The wide range of fatigue ratios shown
by composites relates, in part, to the wide spectrum of materials used to
make them, and to the necessarily broad definition of failure: in
particulate composites, failure means fracture; in fibrous composites it
means major loss of stiffness\textquotedblright }. }

{\normalsize 
\begin{figure}[tbp]
{\normalsize \includegraphics[width=5in]{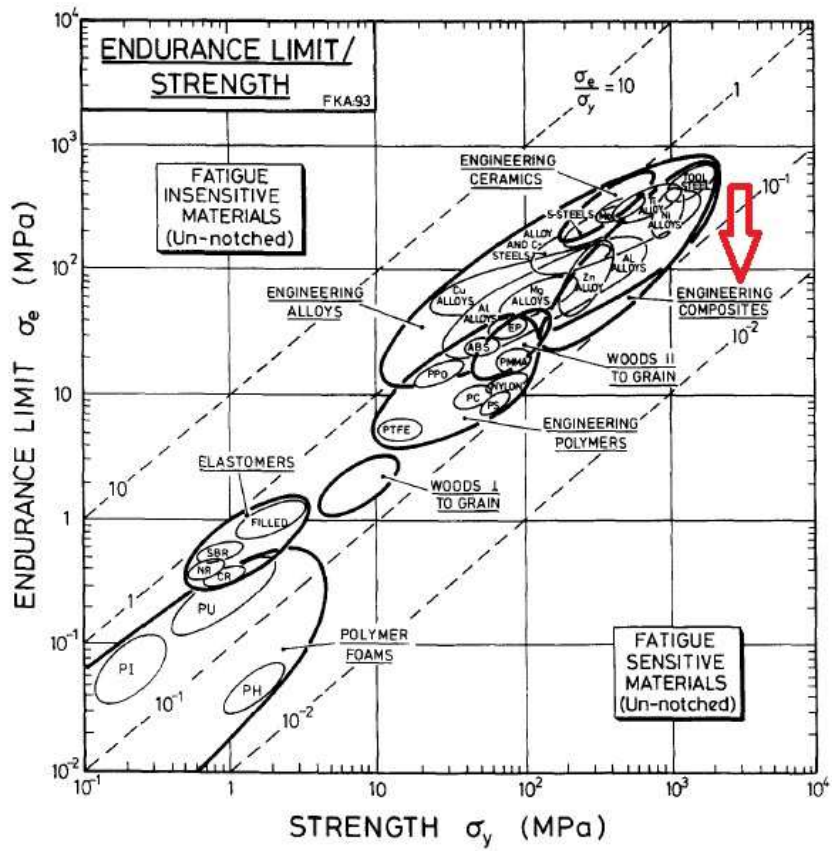} }
\caption{{From Fleck Kang and Ashby [18] (with permission). Classical plot
of static vs fatigue strength for many classes of materials.}}
\label{fig:Fleck}
\end{figure}
}

{\normalsize Clearly, the exact mechanisms for fatigue limit, if there is
one, are microscopic and however may interact with geometry, loading
conditions, etc. For composite materials and structures in general, the
failure mechanism can vary. For materials for which the endurance limit
depends on formation of slip bands, it is obvious to find a correlation with
yield strength, but a full microscopic model for the shape of the SN curve
is more difficult. Fleck et al [] summarize about SN curve: \emph{%
\textquotedblleft It is the failure envelope associated with a sequence of
interdependent phenomena: cyclic hardening, crack nucleation and cyclic
growth, and final fast fracture\textquotedblright }.\ For composites, the
actual nature of each phenomenon is very different, but their
interdependence is also clear. }

{\normalsize K's model is based on some statistical reasoning over the
distribution of static strength, and the successive application of cycles.
Making a certain number of (reasonable) assumption, he seems to derive
apparently simple and clear results, which then he combines with a
calculation of probability of failure. We shall discuss in the present note
the basic results of K model in details. }

\section{\protect\normalsize Main assumptions in K's model}

{\normalsize K's model in the original form [3] makes a certain number of
assumptions, including that the probability of failure stays constant,
cycle-by-cycle. Since the probability of failure during the first cycle is
determined by a probability distribution function for the static strength,
the author concludes that it remains the same for all subsequent cycles. In
the later paper [4] , this was obtained \textquotedblleft
rigorously\textquotedblright . He then has to compute the probability of
failure, and essentially, he consider each cycle as an independent event
without any discussion of this assumption. In fact, even if the probability
of failure did remained constant cycle-per-cycle, this is incorrect
calculation for a fatigue experiment. }

{\normalsize As an illustrative example of this incorrect calculations let
us consider the following discrete analogue: }

\begin{enumerate}
\item {\normalsize There is a bucket full of balls that are numbered with 1
(which is analogous to failure of the specimen under a certain applied
stress, ``success'') and 0 (unsuccessful event, no failure of the specimen).
Assume that the probability of failure is $p$, so that the probability of
no-success is $q=1-p$. }

\item {\normalsize One can perform a series of experiments counting the
number $k$ of trials until the picked ball has a number $1$ (success) on it.
This event is called \textquotedblleft first success\textquotedblright.
After each attempt, in order to implement an analogue of the main assumption
of the K's model, one must put the picked ball back into the bucket: then, 
\emph{and only then} the probability of success is \emph{exactly the same}
at each single trial. 
The resultant distribution of number of trials until ``first success'' is
described by the geometric distribution density function: the probability $%
p_{1}$ that the ``first success" occurs at the $k$th trial (cycle) is%
\begin{equation}
p_{1}\left( k\right) =p\left( 1-p\right) ^{k-1}.
\label{probabilty first success}
\end{equation}%
Eq. (\ref{probabilty first success}) is simply the probability of success
(failure) at the last attempt (cycle) $k$ times the probability of
no-failure at previous $k-1$ attempts (cycles). The average expected number $%
\left\langle n\right\rangle $ of trials until the first success is then
given by 
\begin{equation}
\langle n\rangle =\sum\limits_{k=1}^{+\infty }kp\left( 1-p\right) ^{k-1}=%
\frac{1}{p},  \label{average number of trials}
\end{equation}%
as indeed should be expected. }

\item {\normalsize We can also calculate the probability that the failure of
the sample occurs during the first $n$ experiments (cycles), i.e. that the
event ``first success" happens at any cycle $k$ in between $1$ and $n$. This
probability is the cumulative distribution of 
\eqref{probabilty first
success}%
\begin{eqnarray}
P\left( n\right) &=&\sum\limits_{k=1}^{n}p_{1}\left( k\right)
=\sum\limits_{k=1}^{n}p\left( 1-p\right) ^{k-1}=1-(1-p)^{n}
\label{discrete distribution of trials} \\
&=&1-\exp \left( -n\log \frac{1}{1-p}\right) .  \notag
\end{eqnarray}%
If now the number of trials $n$ is treated as a continuous variable, this
function represents the so-called cumulative exponential distribution with
the mean value%
\begin{equation}
\langle n\rangle =\frac{1}{\log \frac{1}{1-p}}=-\frac{1}{\log (1-p)},
\label{exponential average}
\end{equation}%
and if $p$ is small, the two averages (\ref{average number of trials}) and (%
\ref{exponential average}) become close. }
\end{enumerate}

{\normalsize This trivial example implements the main assumption of K's
model. However, a randomly picked specimen is tested each time, each test is
an independent static test, being the probability of failure $p$ equal to
the probability that the static strength of the specimen is less than the
applied static stress $\sigma $. Obviously, this type of test is irrelevant
to the fatigue phenomenon. \emph{The resulting relation between the mean
number of cycles and the applied load, mistakenly claimed by the author to
be the SN curve, is actually the mean number of tested specimens until
failure!} If the same specimen undergoes subsequent loadings, it may fail
only if its static strength decreases with cycles. However, this static
strength degradation should be a material specific function and is not
uniquely determined by the statistics of the static strength. }

{\normalsize While dispersion in the static strength of a material reflects
the possible level of initial damage observed in the material, the fatigue
failure phenomenon is the result of the initial damage growth and
accumulation with cyclic loading. This growth would eventually appear in any
specimen independently of the strength in other specimens and the statistics
that describes the static strength scatter. }

\section{\protect\normalsize K's model with strength degradation}

{\normalsize In his 2012 PhD thesis [5] and in his 2011 paper [4] the author
extends the model and incorporates the residual strength degradation.
Suppose that a constant amplitude load with maximum stress $\sigma $ ($%
R=\sigma _{min}/\sigma =0$) is applied to a composite structure. If the
static failure strength $\sigma _{fs}$ of this structure is less or equal to 
$\sigma $ (i.e. $\sigma _{fs}\leq \sigma $) the structure will fail at $n=0$%
, i.e. before the first cycle is completed, whilst if $\sigma _{fs}>\sigma $
the structure will fail at the cycle $n=N$. Within the K's wear-out model
the quantity $N\ $is treated as constant and no failure can occur for $0<n<N$%
. These assumptions are, as we show in the sequel, the most critical flaws
of K's model. }

{\normalsize If the test is stopped at any cycle level $n<N$ the structure
would not have failed and would still be able to carry load. However, a
strength test on the structure would show a failure strength $\sigma
_{fs}>\sigma _{r}>\sigma $, where $\sigma _{r}$ is the residual strength.
Hence, during cycling, $\sigma _{r}$ decreases from the static failure
strength $\sigma _{fs}>\sigma $ at $n=0$, to $\sigma $ after $N$ cycles. K's
model starts with the assumption that the change in residual strength is
proportional to the current residual strength, which in the simple case of
zero fatigue limit can be written in the form 
\begin{equation}
\frac{d\sigma _{r}}{dn}=-A\sigma _{r},
\end{equation}%
where $A>0$ is independent of $n$ and $\sigma _{r}$. This wear-out model
assumes that under a fixed amplitude the strength of a specimen with higher
residual strength stress will degrade faster than one with lower residual
strength, which is physically unreasonable. That is why a typical wear-out
model would define the strength degradation rate as a reciprocal to the
current residual strength. The above expression in K's model results in a
residual strength%
\begin{equation}
\sigma _{r}=\sigma _{fs}\left( \frac{\sigma }{\sigma _{fs}}\right)
^{n/\left( N-1\right) }.  \label{residual}
\end{equation}%
Treating $\sigma $ as a constant would results, following K's arguments, in
a Weibull cumulative distribution $P_{W}\left( \sigma _{r};\beta _{r},\alpha
_{r}\right) $ for residual strength with shape and scale parameters%
\begin{equation}
\alpha _{r}=\alpha \frac{N-1}{N-n-1};\qquad \beta _{r}=\sigma ^{n/\left(
N-1\right) }\beta ^{\frac{N-n-1}{N-1}},  \label{final_parameters}
\end{equation}%
which vary with $n$ in a simple manner, implying a clear reduction in
experimental scatter with lower stress level of testing (or longer lives).
In particular, for $n\rightarrow N-1$, $\alpha _{r}\rightarrow \infty $
which means that the distribution converges to the Dirac Delta function, the
residual strength becomes a deterministic function, and $\beta _{r}=\sigma $
consistently to the fact that the residual strength tends exactly to the SN
curve. }

{\normalsize There are many critical inconsistencies that one can spot
immediately: Firstly, $N$ from the SN curve should itself be a variate,
instead of being a deterministic and constant quantity. Secondly, with a
constant amplitude $\sigma $, during cycling the residual strength
distribution cannot approach the SN curve from below, since a specimen that
has strength below $\sigma $ should have failed at an earlier stage and for
any $n$ the residual strength distribution should be truncated from below by
the applied stress $\sigma $. One can even and easily show that, in contrast
with what has been claimed by K, K's model cannot lead to a cycle-by-cycle
constant probability of failure $p\left( n\right) $, which indeed, even
within the hypothesis of K's wear out model, would be 
\begin{equation}
p\left( n\right) =\delta _{0n}P_{W}\left( \sigma ;\beta ,\alpha \right)
+\delta _{nN}\left[ 1-P_{W}\left( \sigma ;\beta ,\alpha \right) \right] ,
\label{probabylity cycle by cycle}
\end{equation}%
where $\delta _{kh}$ is the Kronecker delta (see Appendix II). The discussed
inconsistencies in the model development refute the claim of rigorous proof
of constant cycle-by-cycle probability of failure. 
}

{\normalsize 
}

\section{\protect\normalsize Comparisons with experimental data}

{\normalsize The original K model did not find an exact Weibull distribution
for the life, but over a very wide range of $p$ values ($p<0.1$ hence,
unless the applied load is very high, and close to the static strength) the
two ratios of mean and modal lives to B-Basis life (respectively 17.86 and
8.93)\ are essentially constant. This suggested the author of the original K
model that the average of the two ratios, 13.4 is very close to the value of
13.6 determined in the NAVY reports [6] after statistical analysis of
thousands of data points. \textit{\ } }

{\normalsize Using a slightly different re-derivation (see Appendix I), we
show that K's approach essentially obtains an exponential distribution of
fatigue lives, for a Weibull starting point in scatter of static data.
Hence, it is correct to say that K's approach, in a slightly re-elaborated
form (see Appendix I), we easily obtain the estimate on fatigue ratio (FR,
defined as the ratio between fatigue limit at 10\symbol{94}6 cycle, and the
``static value"\ at $N=1$), as 
\begin{equation}
FR\left( \alpha \right) =\frac{\sigma _{\max ,\lim ,10^{-6}}}{\sigma _{\max
,1}}=10^{\left( -6/\alpha \right) },
\end{equation}%
where $\alpha $ is a Weibull shape parameter in a 2 parameters Weibull
distribution, of the static strength distributions. Notice that in this
form, K's model includes also the R-ratio effect, for $0<R<1$ ie in the
cases of pure tensile loading. For $\alpha $ we can make use not of few
sparse references like K, but the thousands of test done by the Navy. The
distribution of fatigue lives is not as unique as to be a exponential $%
\alpha _{L}=1$. \ A full distribution was found, which we may define the
distribution of ``scatter of fatigue lives'', $\alpha _{L}$. A generally
accepted approximation is to take $\alpha _{L}=1.25$, but it is clear that
its distribution is relatively wide, and depends also on the method used for
analysis. }

{\normalsize 
}

{\normalsize Also, the distribution of static strength scatter $\alpha $ has
in turn a distribution, which we can denominate $\alpha _{a}$, see Figure %
\ref{fig:Navy2}. }

{\normalsize 
\begin{figure}[tbp]
{\normalsize \includegraphics[width=4in]{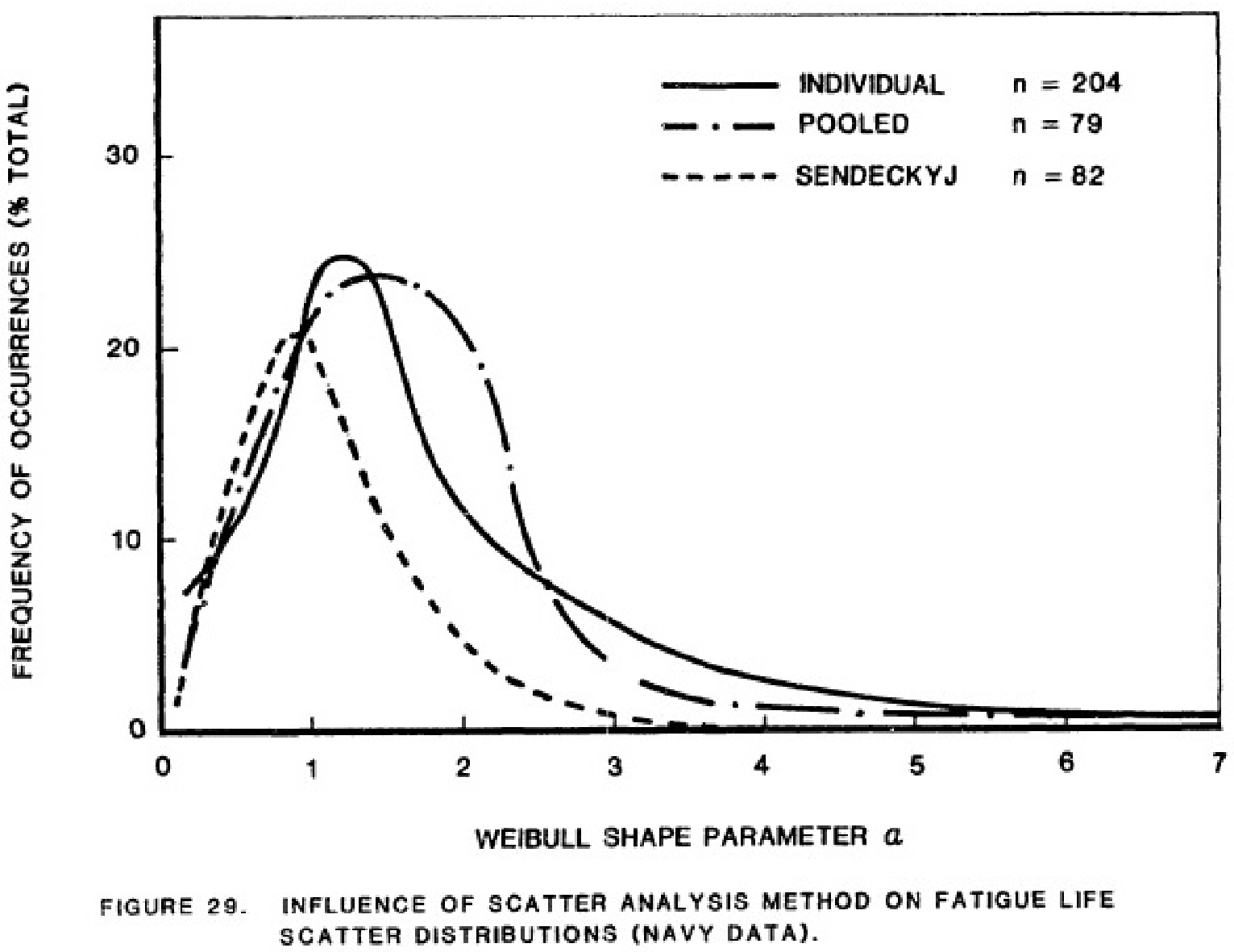} %
\includegraphics[width=4in]{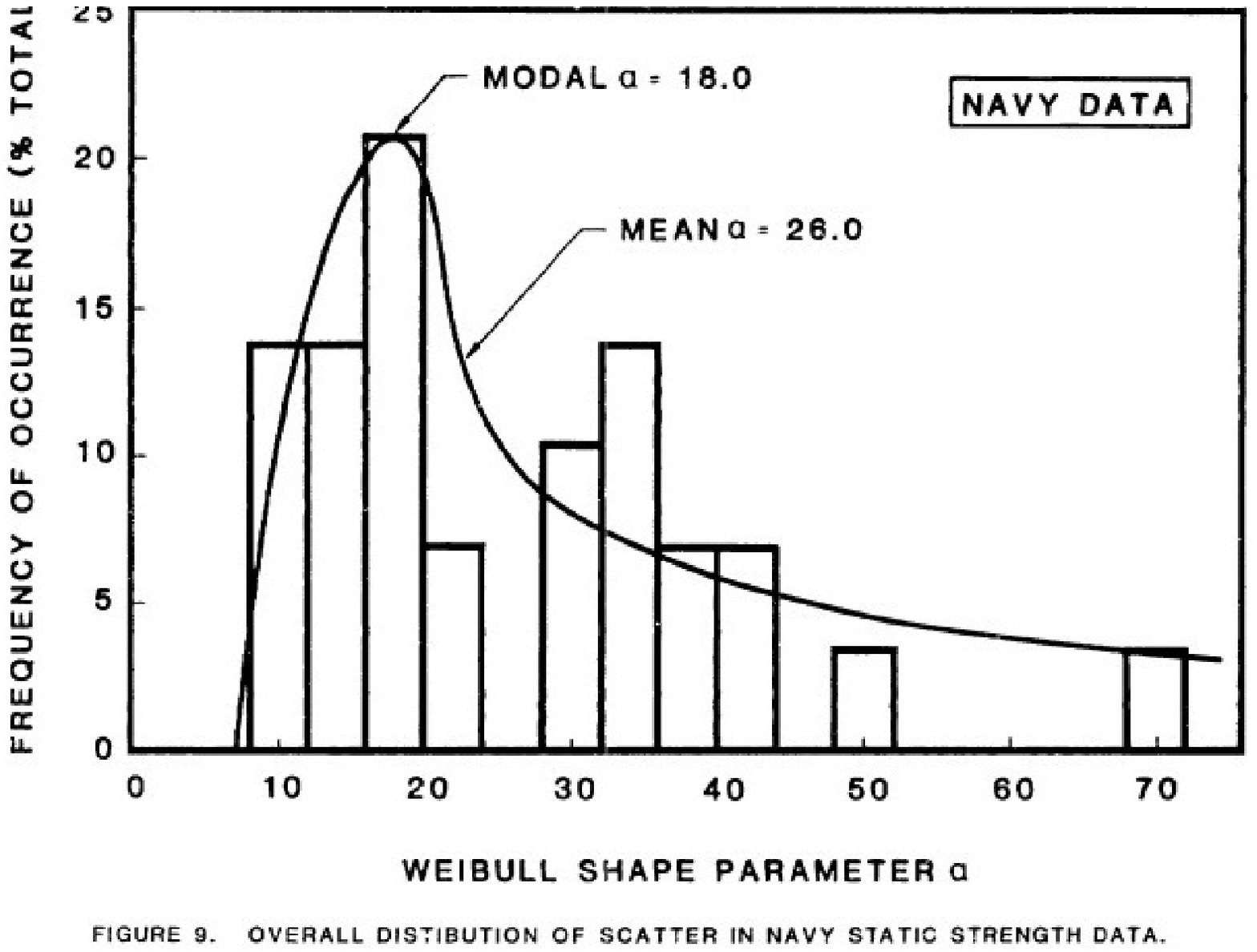} }
\caption{{(a)\ distribution of fatigue lives scatter $\protect\alpha _{L}$
and (b) of scatter of static strength $\protect\alpha $}}
\label{fig:Navy2}
\end{figure}
}

{\normalsize We can use the mean value of $\alpha $ which results from the
Figure \ref{fig:Navy2} to be 26. This corresponds to a value%
\begin{equation}
FR\left( \alpha _{mean}\right) =10^{\left( -6/26\right) }=0.5878,
\end{equation}%
which is outside the known values of Ashby fatigue ratios. This suggests, as
confirmed by most experimental data we shall describe, that K's method would
tend to give unconservative estimates, as too high FR. This however depends
very much in the type of materials under examinations. If the materials are
of "poor" quality, full of defects, tending to having low }$\alpha ,$ 
{\normalsize then the very steep SN curve predicted by K may neglect the
possible phase in SN curve where the degration is not so evident, resulting
extremely conservative. On the other hand, for materials having very high }$%
\alpha ${\normalsize \ like close to a metal, K's theory fails to capture
the wearout at all, and results in too optimistic SN curve. Unidirectional
laminates will tend to have very horizontal fatigue lines, yet their static
scatter may be significant.}

{\normalsize Hence, even if the agreement with Navy experiments has some
very loose qualitative agreement in terms of scatter of fatigue lives, this
is an oversimplification (a single mode value instead of the full
distribution) and the huge risks of using this approximation even as a crude
estimate, is evident.}

{\normalsize The recent investigation on K's method by the FAA (Tomblin and
Seneviratne, 2011 [5], Appendix A) finds also the SN curve predicted by the
original K model (which we found here is the mean life curve) to have rather
erratic comparison with experimental data. In particular, in 14 sets of
data, K's model was found }

\begin{itemize}
\item {\normalsize accurate only for 2 sets }

\item {\normalsize conservative only for 2 sets (both by 1-2 orders of
magnitude) }

\item {\normalsize \textit{unconservative }for the vast majority of data (10
sets), of which 5 perhaps by 1-2 orders of magnitude, 3 by 2-3 orders of
magnitude, and 2 by 4-5 orders of magnitude. }
\end{itemize}

{\normalsize Clearly, although this statistics don't say much, it shows a
tendency of K's method to overpredict fatigue life by large factors.
Examples given in chapt.6 of [3] and in [19] show that more sophisticated
methods with variable $p$ function may improve the situation, although the
examples given tend to predict longer lives than the original K method ---
we are not able to judge if the errors and approximations from the original
K theory with "constant\ p value" continue to manifest their negative effect
here. S/N curves based on the Sendeckyj analysis [2] were found generally
accurate and conservative, but this is to be expected since that method is a
fitting method of SN data. }

{\normalsize In a recent book, Vassilopoulos \& Keller [16] compare 4
methods to make a statistical analysis of fatigue data, which is a problem
of enormous industrial interest since aeronautical structures are designed
and certified using SN curves that correspond to high reliability levels in
the range above 90\% and conform with design codes, but without an
impossibly expensive program of fatigue testing on a population of full
scale \ structures. \ The method based on the normal lifetime distribution
(\textquotedblleft NLD\textquotedblright ) was found non-conservative,
giving a median SN curve which is closer to the median SN curve of ASTM than
the 95\% reliability one. Whitney's pooling scheme and Sendeckyj's wear-out
model are found to produce similar SN curves, with Whitney's easier to
implement, as not requiring any optimization process, and Sendeckyi being
also less conservative. However, this is mainly due to the need of multiple
fatigue results at each stress level, and no capability to consider static
data. Some significant problems were found in the fitting of Sendeckyi's
constant process, with strange slopes of the SN curve predicted,
particularly when disregarding static strength data. A discussion follows on
the appropriateness of including the static data in the fitting. }

{\normalsize K's method is not even compared to the previous four, mainly
because the static data were not enough to fit Weibull distributions. It is
discussed however in its extension to describe mean stress effect, in a
later chapter on Constant Life Diagrams. However, its assumptions are
negatively judged \textit{``This assumption oversimplifies the reality and
masks the effect of the different damage mechanisms that develop under
static loading and at different stages of fatigue loading.'', and ``the
restricted use of static data disregards the different damage mechanisms
that develops during fatigue loading and in many cases leads to erroneous
results''}. In the evaluation of K's model for one database the model 
\textit{``proved to be inaccurate for the examined material's fatigue data''}%
. }

{\normalsize 
%
}

\section{\protect\normalsize Conclusion}

{\normalsize In the original K's model there is a confusion between what we
call fatigue and statistics of the static strength of a number of specimens.
Fatigue life (number of cycles) is mistakenly replaced with the number of
tested specimens to find a specimen with strength less than applied load.
This number of specimen indeed solely depends on the initial statistical
distribution of the static strength, while fatigue is related to damage
accumulation in a specimen and its strength degradation with cycles, which
contradicts the main assumption. One can also mistakenly deduce from the
proposed model that if there is no dispersion in the static strength, for
instance, all the specimen have exactly the same static strength, there is
no such thing as an SN curve. K model is an interesting attempt of using
wear-out models with degradation deterministic equations to predicting SN
curves from static data only for composites (something which is not easy
even with metals). However, its results don't look realistic at all, and
indeed we have explained here why. Not surprisingly, SN curves found in many
independent assessments were found to be (generally) \textit{unconservative }%
for the vast majority of data (10 sets) considered in FAA 2011 report [15],
at least by 1-2 orders of magnitude. We have given reasons for this effect,
both theoretically and with additional estimates from large set of results
from databases of composite materials. }

{\normalsize Only "fitting" models can be considered reliable, as discussed
by Vassilopoulos \& Keller [16], and it should be remarked in this respect
that an additional interesting wearout model is [21-23]. }

\section*{\protect\normalsize Appendix I - SN curve of K's model in slightly
different form}

{\normalsize We have shown that K's model is incorrect. \ However, a simpler
form can be adapted for comparative form in a much simpler form. In
particular, using this equation for a SN curve at any quantile $Q$ }

{\normalsize 
\begin{equation}
N=\left( \frac{\beta }{\sigma }\right) ^{\alpha }\left[ -\log \left(
1-Q\right) \right] ,
\end{equation}%
which obviously has mean value $N_{m}=\left( \beta /\sigma \right) ^{\alpha
} $ and modal value $N_{\text{mod}}=\left( \beta /\sigma \right) ^{\alpha
}\log 2$, but mode value zero (because the distribution of lives is an
exponential distribution $\alpha _{L}=1$), we obtain a closed form version
of the "incorrect" K model, which can be used more easily than the original
K model which is not in closed form, and which obtains only the \textit{mode}
life $N_{c}$ 
\begin{equation}
N_{c}=\left( \frac{\beta }{\sigma }\right) ^{\alpha },
\end{equation}%
which in the present result, coincides with the present mean value. The
distribution in terms of stress for given number of cycles, $P_{W}(\sigma ;%
\frac{\beta }{N^{1/\alpha }},\alpha )$ }

{\normalsize 
\begin{equation}
\sigma =\left( \frac{\beta }{N^{1/\alpha }}\right) \left[ -\log \left(
1-Q\right) \right] ^{1/\alpha };
\end{equation}%
this has obviously mean value $\sigma _{m}=\left( \beta /N^{1/\alpha
}\right) \Gamma \left( 1+\frac{1}{\alpha }\right) $, whereas median value $%
\sigma _{med}=\left( \beta /N^{1/\alpha }\right) \left( \log 2\right)
^{1/\alpha }$. In other words, in this new form, the SN derives from Weibull
distribution both in terms of stress and life at all levels of stress
including the original static distribution $P_{W}(\sigma ;\beta ,\alpha )$
which is obtained consistently for $N^{1/\alpha }=1$. }

\section*{\protect\normalsize Appendix II - Inconsistencies of Kassapoglou's
model}

{\normalsize Suppose we want to calculate the cumulative probability
distribution of failure $P\left( n\right) $ during the first $n$ cycles
assuming K's wear-out model%
\begin{equation}
\frac{\sigma _{r}}{\sigma _{fs}}=\left( \frac{\sigma }{\sigma _{fs}}\right)
^{n/\left( N-1\right) },  \label{wear out model}
\end{equation}%
where the $\sigma _{r}$ is the residual strength, $\sigma _{fs}$ is the
static strength, $\sigma $ is the applied fatigue load, $n$ is the actual
cycle number, and $N$ is the number of cycle at which the samples fails
(assuming its static strength is larger than $\sigma $). Equation (\ref{wear
out model}) simply states that: (i) all sample which have a static strength $%
\sigma _{fs}$ larger than the fatigue stress $\sigma $ will fail at the same
given number of cycles $n=N$, and (ii) samples with static strength $\sigma
_{fs}$ less than $\sigma $ will all fail at cycle $n=0$. Therefore samples
may fail either at $n=0$ when $\sigma _{sf}\leq \sigma $ or at $n=N$ when $%
\sigma _{sf}>\sigma $, no failure may occur in between i.e. for $0<n<N$.
Within the K's wear-out model we have: (a) the probability that failure
occurs at $n=0$ is $p_{0}=P_{W}\left( \sigma ;\beta ,\alpha \right) $, (b)
the probability that failure occurs at $n=N$ is $p_{N}=P\left( \sigma
_{S}>\sigma \right) =1-P_{W}\left( \sigma ;\beta ,\alpha \right) $, (c) the
probability that failure occurs at $n$ satisfying the condition $0<n<N$ is $%
p_{n}=0$. }

{\normalsize Now let us calculate the cumulative probability distribution of 
$P_{n}=\sum_{k=0}^{n}p_{k}$ with $n\geq 0$, which is the probability that
the sample fails within the first $n\geq 0$ cycles, i.e. $P_{n}=P\left(
0\leq k<n\right) $. Since failure cannot occur in between $0$ and $N$ (i.e. $%
p_{k}=0$ for $0<k<N$) the probability that failure occurs within the firs $%
n<N$ must be equal to the probability that failure occurs at cycle $0$, i.e. 
\begin{equation*}
P_{n}=p_{0}=P_{W}\left( \sigma ;\beta ,\alpha \right) ;\qquad 0\leq n<N,
\end{equation*}%
whereas considering that for $n\geq N$ failures has necessarily occurred one
as%
\begin{equation*}
P_{n}=1;\qquad n\geq N.
\end{equation*}%
Therefore the cumulative distribution presents two steps one of amplitude $%
P_{W}\left( \sigma \right) $ at $n=0$ ant the other of amplitude $%
1-P_{W}\left( \sigma ;\beta ,\alpha \right) $ at $n=N$. In between the
cumulative probability distribution is constant. We stress that, as already
shown, the probability of failure cycle per cycle is not constant indeed it
is:%
\begin{align*}
p_{n}& =P_{W}\left( \sigma ;\beta ,\alpha \right) ;\qquad n=0, \\
p_{n}& =0;\qquad 0<n<N, \\
p_{n}& =1-P_{W}\left( \sigma ;\beta ,\alpha \right) ;\qquad n=N, \\
p_{n}& =0;\qquad n>N,
\end{align*}%
in compact notation%
\begin{equation}
p\left( n\right) =\delta _{0n}P_{W}\left( \sigma ;\beta ,\alpha \right)
+\delta _{nN}\left[ 1-P_{W}\left( \sigma ;\beta ,\alpha \right) \right] ,
\label{probability cycle per cycle}
\end{equation}%
where $\delta _{jk}$ is Kronecker's delta. Eq. (\ref{probability cycle per
cycle}) shows that the probability of failure cycle per cycle is zero for $%
0<n<N$, thus revealing one of the serious mistakes of the Kassapoglou model
where the cycle per cycle probablity of failure was assumed different from
zero and equal to $P_{W}\left( \sigma ;\beta ,\alpha \right) $. Indeed, Eq. (%
\ref{probability cycle per cycle}) shows that the sample life $n$ is a
discrete statistical quantity which only takes two different values $n=0$
and $n=N$, and failure at $n=0$ occurs with probability $P_{W}\left( \sigma
;\beta ,\alpha \right) $ wherease failure at $n=N$ occurs with probability $%
1-P_{W}\left( \sigma ;\beta ,\alpha \right) $. This allows to calculate
within the wear-out model (\ref{wear out model}) the expected life of the
samples as 
\begin{equation}
\left\langle n\right\rangle =N\left[ 1-P_{W}\left( \sigma ;\beta ,\alpha
\right) \right] .  \label{mean life}
\end{equation}%
which as expected differs from the value obatined by K. }

\section*{\protect\normalsize References}

{\small \noindent 1. Hahn HT and Kim RY. Proof testing of composite
materials. \textit{J Composite Materials} 1975; 9: 297-311.\vspace{10pt} }

{\small \noindent 2. Sendeckyj GP. Fitting models to composite materials
fatigue data. test methods and design allowables for fibrous composites, In:
Chamis CC (ed) \textit{ASTM STP 734. }Philadelphia, PA: American Society for
Testing and Materials, 1981, 245-260.\vspace{10pt} }

{\small \noindent 3. Kassapoglou C. Fatigue life prediction of composite
structures under constant amplitude loading. \textit{J of Composite
Materials }2007; 41: 2737-2754. 
\vspace{10pt} }

{\small \noindent 4. Kassapoglou C. fatigue model for composites based on
the cycle-by-cycle probability of failure: implications and applications, 
\textit{J of Composite Materials} 2011; 45: 261-277.\vspace{10pt} }

{\small \noindent 5. Kassapoglou C. 2012. \textit{Predicting the structural
performance of composite structures under cyclic loading}, PhD thesis, Delft
Univ of Technology.\newline
http://repository.tudelft.nl/assets/uuid:73a4025d-c519-4e3a-b1cd-c1c8aa0fdfeb/%
\newline
full\_document\_v3.pdf.\vspace{10pt} }

{\small \noindent 6. Juvinall RC and Marshek KM. \textit{Fundamentals of
machine component design}. 5th ed. John Wiley \& Sons Inc, 2011.\vspace{10pt}
}

{\small \noindent 7. Lee J-W, Daniel IM and Yaniv G. Fatigue life prediction
of cross-ply composite laminates. In: Lagace PA (ed) \textit{Composite
Materials: Fatigue and Fracture, Second Volume. ASTM STP 1012.}
Philadelphia, PA: American Society for Testing and Materials, 1989, pp.
19-28.\vspace{10pt} }

{\small \noindent 8. Gathercole N, Reiter H, Adam T and Harris B. Life
prediction for fatigue of T800/5245 carbon-fibre composites: I. Constant
amplitude loading. \textit{Fatigue} 1994; 16: 523-532.\vspace{10pt} }

{\small \noindent 9. Amijima S, Fujii T and Hamaguchi M. Static and fatigue
tests of a woven glass fabric composite under biaxial tension-torsion. 
\textit{Composites }1991; 22: 281-289.\vspace{10pt} }

{\small \noindent 10. Cvitkovich MK, O'Brien TK and Minguet PJ. Fatigue
debonding characterization in composite skin/stringer configurations. In:
Cucinell RB (ed), \textit{ASTM STP 1330}, Philadelphia, PA: American Society
for Testing and Materials, 1998, pp. 97--121.\vspace{10pt} }

{\small \noindent 11. O'Brien TK. Fatigue delamination behavior of peek
thermoplastic composite laminates. \textit{J. Reinforced Plastics and
Composites} 1988; 7: 341-359.\vspace{10pt} }

{\small \noindent 12. O'Brien, TK, Rigamonti M and Zanotti C. \textit{%
Tension fatigue analysis and life prediction for composite laminates}.
Hampton, VA: National Aeronautics and Space Administration. Technical
Memorandum 100549, 1988.\vspace{10pt} }

{\small \noindent 13. Maier G, Ott H, Protzner A and Protz B. Damage
development in carbon fibre-reinforced polyimides in fatigue loading as a
function of stress ratio. \textit{Composites }1986; 17: 111--120.\vspace{10pt%
} }

{\small \noindent 14. Gerharz JJ, Rott D and Schuetz D.
Schwingfestigkeitsuntersuchungen an Fuegungen in Faserbauweise, \textit{%
BMVg-FBWT}, 1979. pp. 79-23.\vspace{10pt} }

{\small \noindent 15. Tomblin J and Seneviratne W. \textit{Determining the
fatigue life of composite aircraft structures using life and
load-enhancement factors}. Report DOT/FAA/AR-10/6. Federal Aviation
Administration, National Technical Information Service, Springfield, VA.
2011. www.tc.faa.gov/its/worldpac/techrpt/ar10-6.pdf\vspace{10pt} }

{\small \noindent 16. Vassilopoulos AP and Keller T. \textit{Fatigue of
fiber-reinforced composites}, London: Springer-Verlag, 2011. \vspace{10pt} }

{\small \noindent 17. Andersons J and Paramonov Yu. Applicability of
empirical models for evaluation of stress ratio effect on the durability of
fiber-reinforced creep rupture-susceptible composites. \textit{J Mater Sci}
2011; 46: 1705-1713. 
\vspace{10pt} }

{\small \noindent 18. Fleck NA, Kang KJ, Ashby MF. Overview no. 112: The
cyclic properties of engineering materials, \textit{Acta Metallurgica et
Materialia } 1994; 42: 365--381 }

{\small \noindent 19. Whitehead RS, Kan HP, Cordero R and Saether ES. 
\textit{Certification Testing Methodology for Composite Structures, Vol I
and II}. 
Naval Air Development Centre Report No. 87042-60 (DOT/FAA/CT-86-39), 1986.%
\newline
http://www.dtic.mil/dtic/tr/fulltext/u2/b112288.pdf\vspace{10pt} }

{\small \noindent 20. Kassapoglou C and Kaminski M. Modeling damage and load
redistribution in composites under tension-tension fatigue loading, \textit{%
Composites: A} 2011; 42: 1783--1792.\vspace{10pt} }

{\small \noindent 21. D'Amore A, Caprino G, Stupak P, Zhou J and Nicolais L.
Effect of stress ratio on the flexural fatigue behaviour of continuous
strand mat reinforced plastics. \textit{Science and Engineering of Composite
Materials} 1996; 5: 1-8.\vspace{10pt} }

{\small \noindent 22. Caprino G and D'Amore A. Flexural fatigue behaviour of
random continuous fibre reinforced thermoplastic composites. \textit{%
Composite Science and Technology} 1998; 58: 957-965.\vspace{10pt} }

{\small \noindent 23. D'Amore A, Caprino G, Nicolais L and Marino G.
Long-term behaviour of PEI and PEI-based composites subjected to physical
aging. \textit{Composites Science and Technology} 1999; 59: 1993-200. }

\end{document}